\documentclass[%
 reprint,
amsmath,amssymb,
]{revtex4-2}

\usepackage{graphicx}
\usepackage{dcolumn}
\usepackage{bm}
\usepackage{natbib}

\begin{document}


\title{Suppressing convection strength using confinement during protein crystallization}

\author{Tapan Kumar Pradhan}
\email{tapan.k.pradhan@gmail.com}
 \affiliation{Department of Mechanical Engineering, \\ Indian Institute of Technology Kharagpur, Kharagpur 721302, India.} 
 
\author{Pradipta Kumar Panigrahi}%
 
\affiliation{Department of Mechanical Engineering, \\ Indian Institute of Technology Kanpur, Kanpur 208016, India.}


\begin{abstract}
Fluid convection during protein crystallization plays a significant role in determining crystal growth rate and crystal quality. Crystals grown in reduced flow strength gives better quality crystal. Hence, tuning the flow strength is very essential in the crystal growth process. In this work, we have demonstrated a new method to suppress the flow strength using confinement effect during vapor diffusion method of protein crystallization where the crystal is grown inside an evaporating droplet. Flow study is carried out at four different confinement conditions to study the effect of confinement. Flow inside the droplet is caused by evaporation induced natural convection which is measured by micro-PIV method. Also concentration gradient generated around the growing crystal also induces buoyancy driven flow around the protein crystal during the crystal growth phase. The evaporation rate from the droplet and the flow strength inside the droplet get suppressed by increasing confinement. Hence, the flow strength can be tuned by adjusting the confinement which is a very simple method to tune the flow strength.
\end{abstract}

\maketitle


\section{Introduction}
Protein crystallization plays a vital role in structural biology, protein purification, drug delivery etc. Better quality crystal is required for X-ray diffraction study of of protein molecules which is essential for understanding the function of the protein. Many factors affect the crystallization process. Fluid convection in the solution where the crystal growth occurs significantly affects the crystal growth rate and crystal quality \cite{Vekilov1998a,Vekilov1998b,Parambil2011,Roberts2010}. Hence, the fluid dynamics study during the protein crystallization process has been an important subject of research by many researchers \cite{Savino1996,Kawaji2003,Pradhan2012}. Evaporation induced natural convection and buoyancy driven convection due to depletion of protein concentration surrounding the crystal are the two mechanism of flow observed during the crystal growth process \cite{Savino1996,Lin1995}. Numerically studies by \citet{Lin1995,Savino1996} show that buoyancy driven convection is the dominant mode of solute transport during the protein crystal growth process. Experimentally the fluid convection inside the droplet during sitting drop method of protein crystallization was reported by \citet{Pradhan2012}.

\par
Higher convective flow during protein crystal growth adversely affect the crystal quality. Higher fluid convection increases the crystal growth rate due to increase in solute supply to the interface, but higher convection leads to incorporation of impurities on the crystal surface \cite{Vekilov1998a}. Strength of flow velocity affects the the crystal size and number of crystals \citet{Roberts2010,Parambil2011}. Increase in flow strength increases the nucleation and crystal number. Lower fluid convection yields better crystal quality and lower impurity incorporation.  Crystal grown in micro-gravity condition gives better quality crystal due to reduced convective flow \cite{Snell2005}. Lots of experiments are conducted in space environment to grow the protein crystal in reduced gravity \cite{Delucas1989,Strong1992}. Lots of effort have been done to manipulate the flow specially suppressing the convective flow to obtain better quality crystal. Different external forces have been applied to control the fluid convection. \citet{Penkova2005} observed that solution stirring caused by applied electric field enhanced the protein crystal nucleation.  Magnetic field has been used by many researchers to control the gravity which ultimately reduce the buoyancy driven convection during protein crystallization \cite{Heijna2007,Okada2011,Okada2012,Okada2013,Yin2015}. The fluid convection can be damped, stopped, and even reversed by applying a magnetic field \cite{Heijna2007}. \citet{Nakamura2012} studied the protein crystal growth in a high magnetic field gradient and obtained a enhanced quality crystal. \citet{Adawy2013} and \citet{Poodt2009} suggested a method to obtain high quality crystal grown in convection free environment. They have grown the crystal at the ceiling of the growth cell in batch crystallization which gives diffusion limited conditions for the crystal growth. In their method, gravity suppresses the convection than creating buoyancy driven natural convection. Hence, reducing flow strength during protein crystallization always remain an important aspect of many researchers.

\par
In this paper, we demonstrated a simple method to modify the fluid flow strength by varying the confinement condition during protein crystallization. The study was carried out for vapor diffusion method of protein crystallization where the crystals are grown inside an evaporating droplet sitting on a glass surface. In vapor diffusion method, evaporation from the droplet leads to supersaturation of protein leading to crystallization. Detail measurement of flow velocity was carried out inside the droplet to understand the variation of flow dynamics at different confinement condition. Velocity measurement of fluid during the crystal growth process was studied using micro-PIV technique to understand the detail hydrodynamics of the process.

\section{Experimental Details}
\begin{figure}[htb!]
\begin{center}
\includegraphics[width=0.45\textwidth]{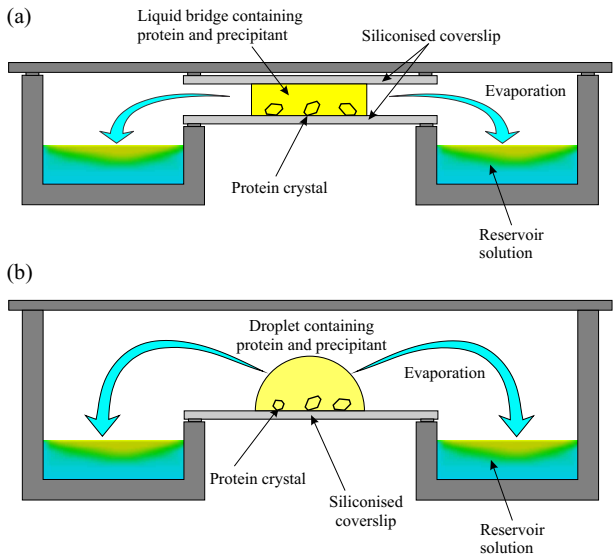}
\caption{\label{fig:chamber}Schematics of crystal growth chamber for (a) sandwich drop method of protein crystallization and (b) sitting drop method of protein crystallization.}
\end{center}
\end{figure}

Sitting drop method of protein crystallization was used in the experiment. In this method, the droplet is placed on a hydrophobic glass cover slip and the cover slip is surrounded by a reservoir solution as shown in Figure \ref{fig:chamber}(b). The confinement effect was studied by squeezing the droplet between two hydrophobic cover slip as shown in Figure \ref{fig:chamber}(a). The droplet forms a liquid bridge between the two cover slips. The two cover slips act as confinement for the evaporation process from the droplet. This method of crystallization is called sandwich method. The confinement condition of protein crystallization is adjusted by adjusting the height of the ceiling (top cover) from the bottom cover slip. 

\par
The protein solution was prepared by dissolving lysozyme protein in de-ionised water. The concentration of the protein solution was kept equal to 70 mg/ml. The lysozyme protein used here is in powder form obtained from Sigma Aldrich. HPLC-grade water was used to prepare the solution.  A reservoir solution was prepared by mixing 250 $\mu$L of aqueous NaCl solution (30 $\%$ W/V), 100  $\mu$L buffer solution (0.1 M aqueous sodium acetate solution having pH of 4.5), 428 $\mu$L aqueous ethyl glycol solution (70 $\%$ W/V) and 222 $\mu$L water. A droplet of volume 2 $\mu$L containing protein solution and reservoir solution in the ratio of 7:3 was placed between two siliconised glass cover slip. The siliconised glass cover slip behaves as a hydrophobic surface. The droplet forms a liquid bridge between the two siliconised glass surfaces. The height of the liquid bridge is approximately equal to 500 $\mu$m. The liquid bridge is of cylindrical in shape having diameter 2.25 mm. The liquid bridge formed between the two glass surfaces is kept inside a crystallization chamber containing the reservoir solution (Figure \ref{fig:chamber}(a)). The results of the sandwich drop method of protein crystallization is compared with the sitting drop method of protein crystallization (Figure \ref{fig:chamber}(b)) to understand the effect of confinement on flow strength. The volume of the droplet is kept same as that of sandwich drop method of protein crystallization. The droplet forms a contact line diameter of 2.03 mm. In case of sitting drop method of protein crystallization, the upper substrate surface does not touch the droplet. Two different configurations for sitting drop method and two configurations for sandwich drop method of protein crystallization was used to study the effect of confinement on flow dynamics. The confinement condition is varied by changing the separation distance between the two substrate surfaces. The inside environment of the crystallization chamber is completely isolated from surrounding room environment. The crystallization is carried out in the room temperature of 20 $^0$C and relative humidity of \(60 \; \% \pm 5 \; \% \). 

\begin{figure}[htb!]
\begin{center}
\includegraphics[width=0.45\textwidth]{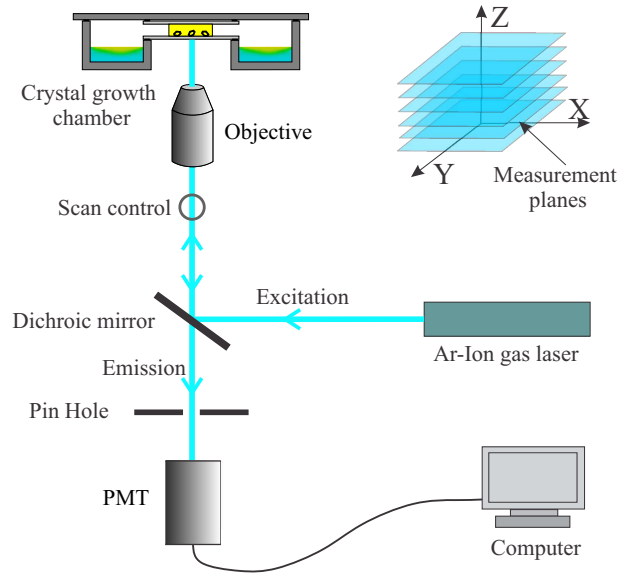}
\caption{\label{fig:exp1}Experimental arrangement to measure the velocity inside the liquid bridge.}
\end{center}
\end{figure}

\par
The crystal growth chamber was designed in such a manner that imaging inside the droplet can easily be done using a confocal microscope. The crystal growth chamber was kept on the stage of the confocal microscope. The experimental setup is shown in Figure \ref{fig:exp1}. Imaging inside the droplet was carried out through the bottom glass surface as shown in Figure \ref{fig:exp1}. The velocity of fluid inside the droplet was measured by micro-PIV technique. Spherical polystyrene particles tagged with fluorescent (Invitrogen Ltd) of diameter equal to 2 $\mu$m were added to the solution of the droplet as tracer particles during PIV measurement. The seeding concentration of these tracer particles was kept at 0.02 $\%$ V/V. The presence of seeding particle in the solution has minimal effect on the protein crystallization \cite{Pradhan2012}. The fluorescent seeding particles are excited using a laser light of wave length 488 nm from Ar-Ion gas laser present in the confocal microscope. The emission from the tracer particles are captured by the PMT present in the confocal microscope. Image is constructed by point scanning of the object by a scanner in the confocal microscope. The images were captured at a time interval of 3 sec. The size of each image was set equal to \(512 \times 512 \) pixels and each image has a field of view equal to \( 2.39 \; \textrm{mm} \times 2.39 \; \textrm{mm}\).

\par
The image processing for the PIV measurement was carried out using Dynamic Studio V1.45 software. Adoptive cross-correlation was used for the PIV evaluation keeping interrogation area equal to \(32 \times 32\) pixels with 25 $\%$ overlap. The resolution of the vector field obtained from the PIV evaluation was equal to \(114 \; \mu \textrm{m} \times 114 \; \mu \textrm{m}\). The velocity measurements were carried out at X-Y plane parallel to the substrate surface. Similar velocity measurements were performed at ten X-Y planes at different Z locations. The velocity vector field obtained from PIV measurements were averaged over 5 measurements to minimize the error caused by Brownian motion \cite{Santiago1998,Meinhart1999}. Velocity of the fluid inside the droplet sandwiched between two cover slips was carried out at different time instants of protein crystal growth to study the dynamic nature of the flow pattern at different stage of crystal growth. Effect of confinement on evaporation induced flow was studied for different geometries during the initial stage of the crystal growth when the flow is dominated by evaporation induced natural convection.

\begin{figure}
\begin{center}
\includegraphics[width=0.4\textwidth]{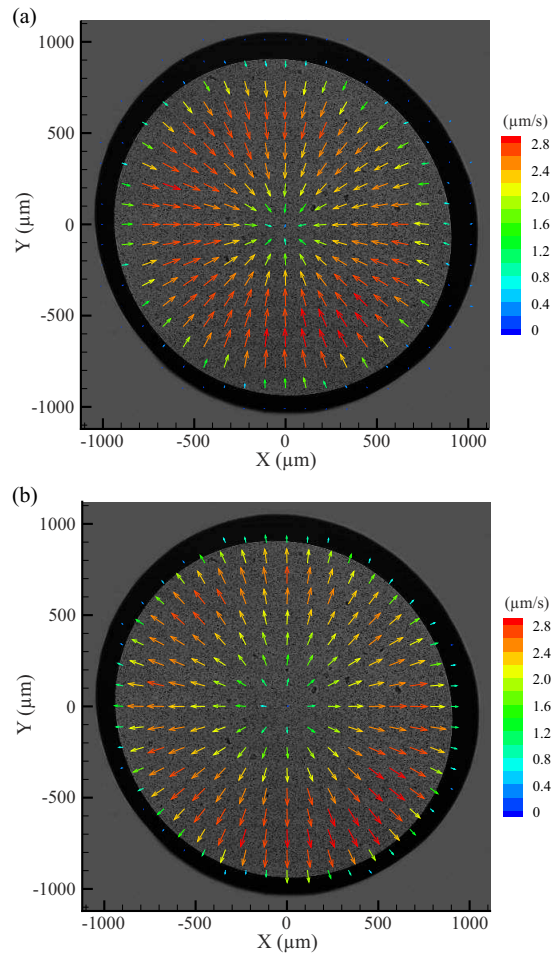}
\caption{\label{fig:vel10}Velocity vector field inside the liquid bridge after 10 minutes of initiation of crystallization process at (a) Z=100 $\mu$m and (b) Z= 400 $\mu$m.}
\end{center}
\end{figure}

\section{Results and Discussion}
The effect of confinement on hydrodynamics of droplet during protein crystallization has been discussed. Then the flow strength inside the droplet is compared at different confinement conditions. Details of the hydrodynamics during protein crystallization are presented in subsequent sections.

\begin{figure}[htb!]
\begin{center}
\includegraphics[width=0.4\textwidth]{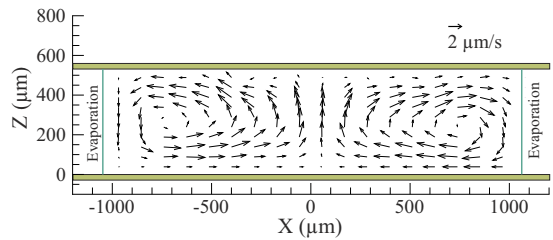}
\caption{\label{fig:3d} Reconstructed velocity vector field in X-Z plane after 10 min of starting of crystallization process at Y=0.}
\end{center}
\end{figure}

\subsection{Hydrodynamics of droplet using confinement}

Fluid flow inside the droplet during protein crystallization in the presence of confinement was presented where the droplet is confined between two parallel cover slips. Velocity vector fields at different time instants are presented subsequently. Figure \ref{fig:vel10} presents the velocity vector field in XY plane obtained from PIV measurement after 10 minutes of starting of the crystallization process. Velocity vector field is shown at two different Z-locations. Figure \ref{fig:vel10}(a) shows velocity vector field near the lower surface at a distance of 100 $\mu$m from the lower substrate surface. The vector field in this plane shows inward flow towards the center of liquid bridge. Figure \ref{fig:vel10}(b) shows velocity vector field near the upper surface at a distance of 400 $\mu$m from the lower surface and the flow shows outward movement from the center. Flow visualization inside the liquid bridge after 10 min is shown in supplementary movie 1. Similar measurements were carried out in 10 X-Y planes at different Z locations. All these X-Y plane velocity measurements only gives $u$ and $v$ velocities at these planes. The complete three-dimensional velocity field is reconstructed using continuity equation (equation \ref{eq:continuity}) and 2D velocity measurements in all these 10 X-Y planes \cite{Pradhan2015a}.

\begin{equation}\label{eq:continuity}
\frac{\partial u}{\partial x}+\frac{\partial v}{\partial y}+\frac{\partial w}{\partial z}=0 
\end{equation}
 
\begin{figure}[htb!]
\begin{center}
\includegraphics[width=0.4\textwidth]{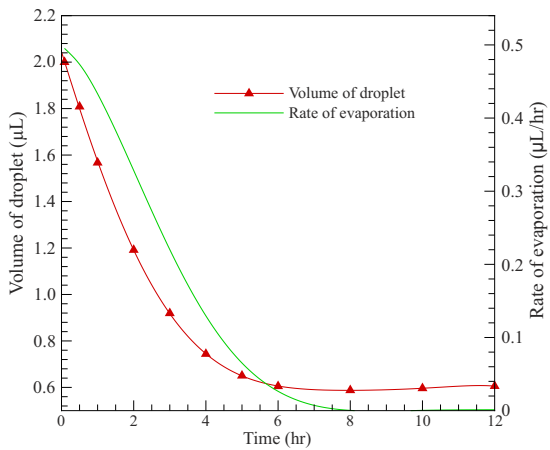}
\caption{\label{fig:evaporation} Volume of liquid bridge and evaporation rate with time.}
\end{center}
\end{figure}

\par
The reconstructed velocity vector field is shown in Figure \ref{fig:3d} along a vertical plane. It shows a downward movement of fluid at the periphery of liquid bridge and an upward movement along the center of liquid bridge causing two circulating loops. The flow is similar to the internal convection of an evaporating liquid bridge of aqueous NaCl solution reported by \citet{Lee2014}. They reported that evaporating liquid bridge of NaCl solution shows buoyancy driven Rayleigh convection inside the liquid bridge. Experimental and numerical studies by \citet{Kang2013,Pradhan2016a} show that convection inside an evaporating droplet of aqueous solution is due to Rayleigh convection. \citet{Lee2014} reported that thermal convection and Marangoni convection is absent in the evaporating liquid bridge. Flow pattern obtained from our experiment and the flow pattern observed by \citet{Lee2014} in liquid bridge of NaCl solution shows similar pattern. Hence, it is expected that the flow pattern observed here is caused by the buoyancy driven Rayleigh convection. Evaporation from the liquid bridge to the reservoir solution occurs due to the vapor pressure difference between the liquid bridge and the reservoir solution. The vapor pressure difference between the liquid bridge and the reservoir solution is because of the different concentration level of the solutions. Evaporation from the liquid-air interface of the liquid bridge increases solute concentration at the interface. Higher solute concentration at the evaporating surface slides down and the lighter fluid at the center of the liquid bridge rises up due to buoyancy creating circulating loop which is shown in Figure \ref{fig:3d}.

\begin{figure}[htb!]
\begin{center}
\includegraphics[width=0.4\textwidth]{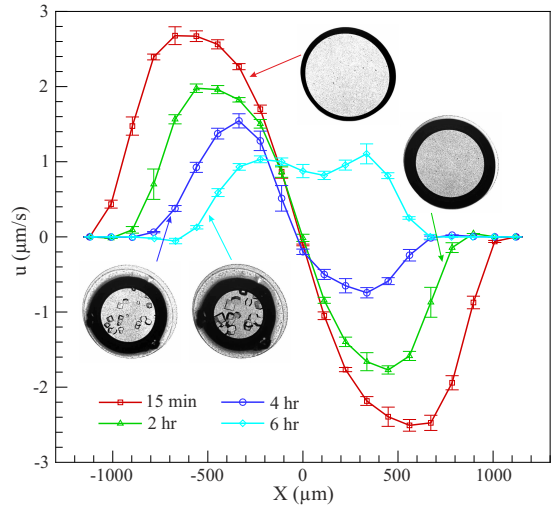}
\caption{\label{fig:profile} X-component of velocity ($u$) at Z=100 and Y=0 at different time of crystallization process.}
\end{center}
\end{figure}

\begin{figure*}
\begin{center}
\includegraphics[width=0.95\textwidth]{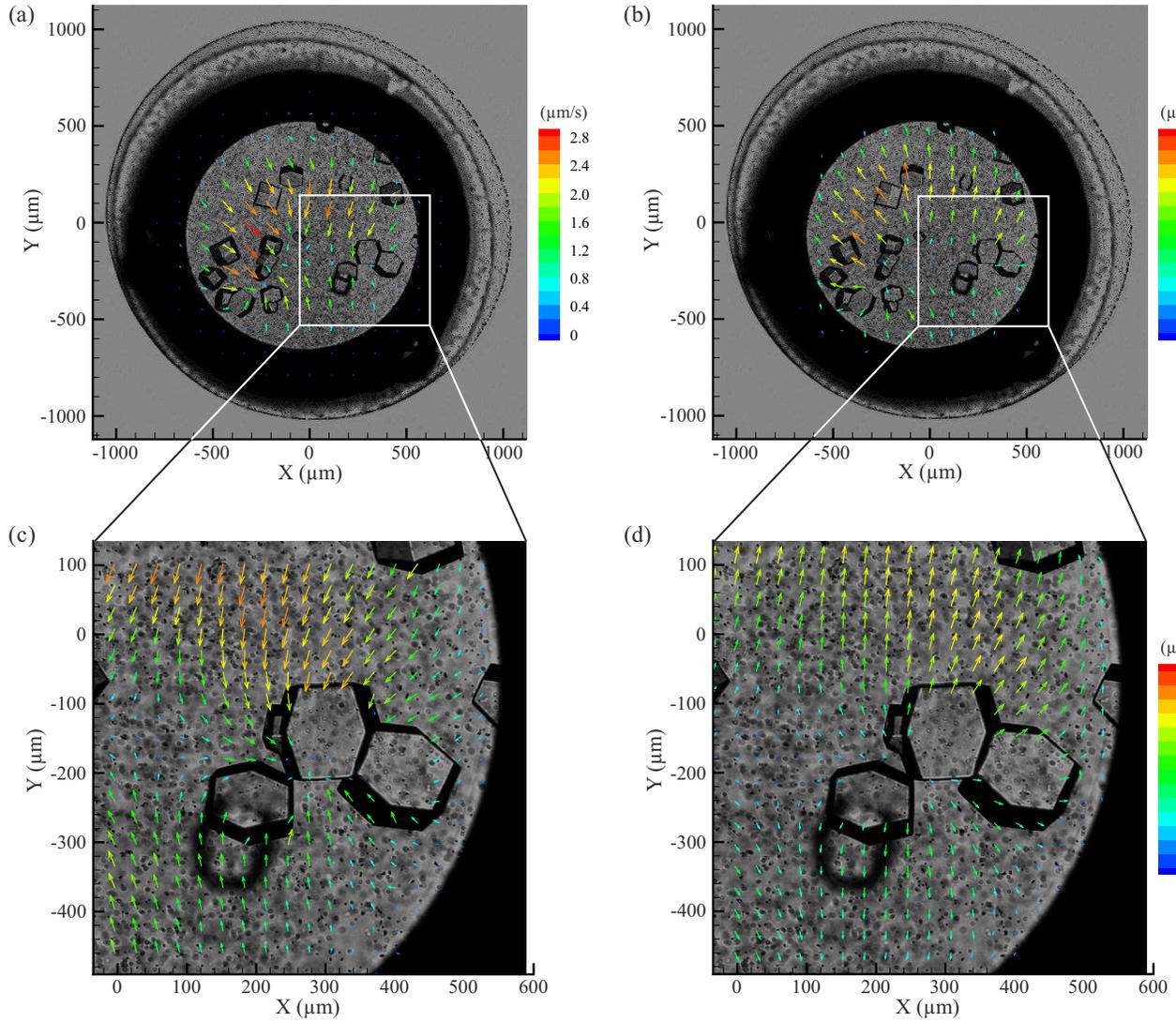}
\caption{\label{fig:velcrystal} Velocity vector field inside the liquid bridge after 4 hr of initiation of crystallization process at (a) Z=100 $\mu$m and (b) Z= 400 $\mu$m.}
\end{center}
\end{figure*}

\par
The volume of the droplet is shown in Figure \ref{fig:evaporation} at different time instant. The volume of the droplet decreases with time and attains constant value after 8 hr. Evaporation rate from the droplet is calculated by taking the slope of the volume curve. The evaporation rate shown in Figure \ref{fig:evaporation} decreases with time. All the transport of water vapor occurs between the droplet and the reservoir solution as the inside environment of the crystal growth chamber is isolated from the outside environment. Evaporation from the liquid bridge increases the solute concentration of the liquid bridge. Hence, vapor pressure of the solution of the liquid bridge get reduced due to increase in solute concentration according to Raoult's law. The vapor pressure difference between the liquid bridge and the reservoir solution which is the driving force for evaporation decreases with time. Hence, evaporation from the liquid bridge reduces with time. Velocity measurements are taken at different time instants during the crystallization process. The $u$-velocity along the diameter (Y=0) of the liquid bridge at z=100 $\mu$m from the substrate surface at different time is shown in Figure \ref{fig:profile}. It shows a reduction in flow strength with time.  Decrease in evaporation rate (Figure \ref{fig:evaporation}) leads to decreases in flow strength as observed in Figure \ref{fig:profile}.

\par
Evaporation from the liquid bridge causes inward movement of the liquid-air interface due to shrinkage caused by water loss. The evaporation induced interface velocity can be given by \(u_i=J/\rho\). Here, $J$ is the evaporation rate and $\rho$ is the density of water. The interface velocity at initial time period is equal to \(3.88 \times 10^{-8} \, \textrm{m/s}\). This value is very less as compared to the observed value from the experiment. Hence, the effect of interface movement on the observed flow is negligible.

\par
Evaporation from the liquid bridge increases the protein concentration leading to super saturation inside the liquid bridge. Combined effect of super saturation and the precipitants leads to nucleation and crystal growth inside the liquid bridge. Crystals start appearing inside the liquid bridge after 3 hours of initiation of crystallization process. When the crystal size increases, the flow pattern inside the liquid bridge gets disturbed due to the presence of crystals. The velocity vector after 4 hours in the presence of crystals is presented in Figure \ref{fig:velcrystal} at 100 $\mu$m and 400 $\mu$m from the lower substrate surface. The flow strength after 4 hours is less compared to the flow strength at 10 min (Figure \ref{fig:vel10}). This is because of the low evaporation rate at later time. Velocity field around a growing crystal after 4 hr at higher magnification is shown in Figure \ref{fig:velcrystal}(c) and (d). Velocity vector field shown in Figure \ref{fig:velcrystal}(c) shows that the fluid moves towards the growing crystals and the velocity vector field shown in Figure \ref{fig:velcrystal}(d) shows the fluid moves away from the protein crystals. The velocity vector fields shows that evaporation induced flow is almost absent. Figure \ref{fig:velcrystal}(c) and (d) show that the fluid moves towards the crystal near the bottom surface and the fluid moves away from the crystal near the upper surface. This flow pattern near the crystals indicates a rising plume like flow pattern near the crystal. Such type of flow near the growing crystal is caused by buoyancy force due to concentration gradient near the crystal \cite{Shlichta1986,Mcpherson1999,Rosenberger1995}. Growing protein crystal causes a depletion zone near the crystal creating concentration gradient. Concentration gradient around the crystal induces buoyancy driven Rayleigh convection. Both evaporation and crystal growth causes flow in post crystal growth phase. The flow pattern after 4 hr (Figure \ref{fig:velcrystal})  is dominated by buoyancy driven natural convection around the growing crystal. The flow visualization inside the liquid bridge after 4 hr in the presence of protein crystals is shown in supplementary movie 2. 

\begin{figure}
\begin{center}
\includegraphics[width=0.4\textwidth]{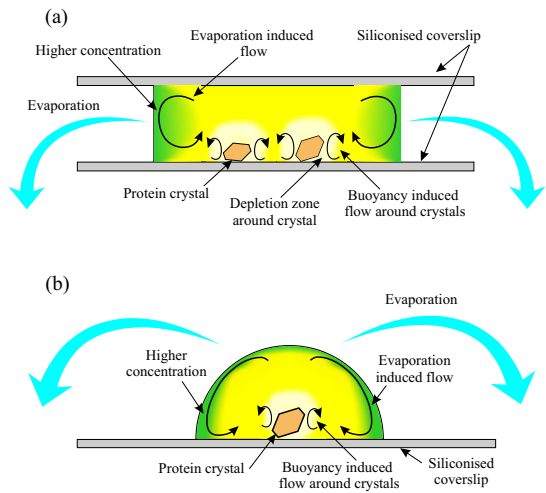}
\caption{\label{fig:depiction} Depiction of fluid convection inside the droplet during protein crystallization for (a) sandwich drop method of protein crystallization and (b) sitting drop method of protein crystallization.}
\end{center}
\end{figure}

\par
The physics of flow pattern inside the liquid bridge during crystal growth process is depicted in Figure \ref{fig:depiction}(a). The flow inside the liquid bridge is induced by both evaporation and crystal growth. During the initial period of crystallization process in the absence of protein crystal, the fluid flow occurs due to evaporation induced Rayleigh convection. With time, evaporation induced convection reduces due to reduction in evaporation rate. When the protein crystals increases to a sufficient size, the flow is dominated by the natural convection induced around the growing crystals. Similar  flow is also depicted in Figure \ref{fig:depiction}(b) for sitting drop method of crystallization.

\begin{figure}
\begin{center}
\includegraphics[width=0.4\textwidth]{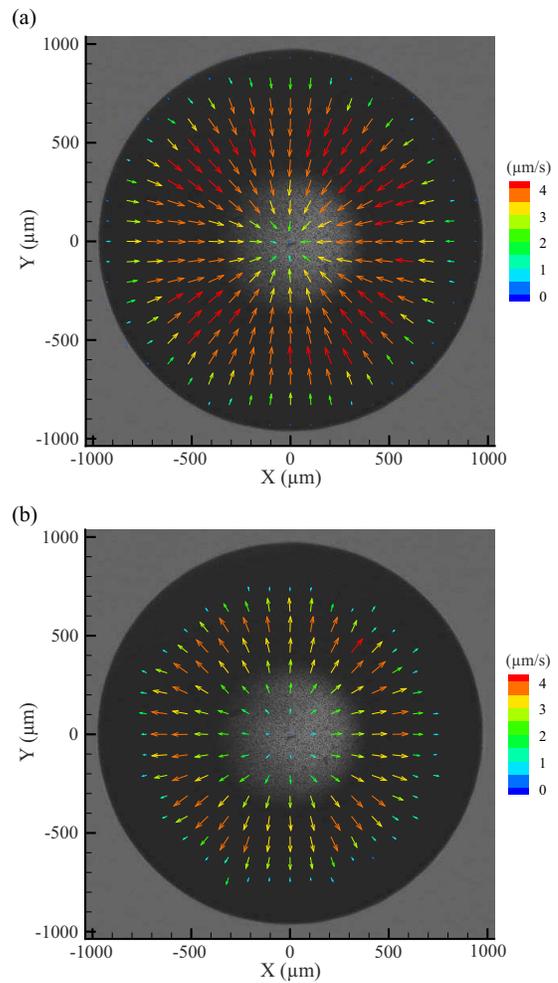}
\caption{\label{fig:veldrop} Velocity vector field inside the droplet during sitting drop method of protein crystallization after 10 min of initiation of crystallization process (a) Z=100 $\mu$m and (b) Z= 450 $\mu$m.}
\end{center}
\end{figure}

\subsection{Comparison at different confinement conditions}

\par
\begin{figure}[htb!]
\begin{center}
\includegraphics[width=0.4\textwidth]{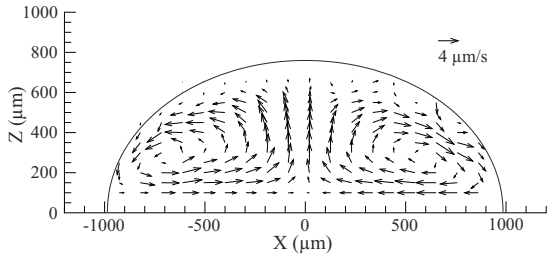}
\caption{\label{fig:3ddrop} Reconstructed velocity field inside the drop during sitting drop method of protein crystallization after 10 min at Y=0.}
\end{center}
\end{figure}

The flow strength of liquid bridge during sandwich method of protein crystallization (using confinement as shown in Figure \ref{fig:chamber}(a)) is compared with the flow strength inside droplet during sitting drop method of protein crystallization without confinement (Figure \ref{fig:chamber}(b)). The volume of droplet in both the case is kept equal to 2 $\mu$L. The velocity vector field inside the droplet of sitting drop method of protein crystallization is presented in Figure \ref{fig:veldrop} at 100 $\mu$m and 450 $\mu$m from the substrate surface after 10 min of initiation of crystallization process. The flow pattern shows inward flow along the substrate surface and outward flow at the apex of droplet. The flow pattern observed here shows similar pattern as that of liquid bridge in case of sandwich method of protein crystallization. The reconstructed velocity vector field is presented in Figure \ref{fig:3ddrop}. It shows a recirculating flow pattern which is same as that of sandwich drop method of protein crystallization. The flow pattern observed inside the droplet is same as that of the flow pattern observed inside evaporating droplet of aqueous NaCl solution \cite{Pradhan2016a,Kang2013}. The fluid convection inside aqueous droplet is only caused by buoyancy driven Rayleigh convection. Hence, the flow pattern observed here is because of Rayleigh convection as presented in Figure \ref{fig:depiction}(b).  

\par
\begin{figure}[htb!]
\begin{center}
\includegraphics[width=0.4\textwidth]{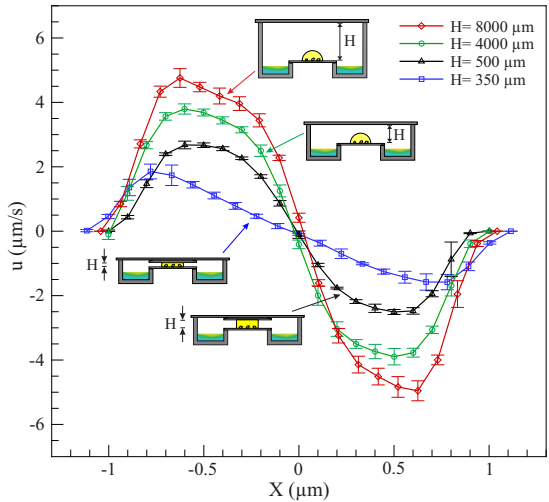}
\caption{\label{fig:profiledrop} Comparison of flow strength inside droplet at different confinement conditions.}
\end{center}
\end{figure}

The flow strength for sitting drop method and sandwich method of protein crystallization is compared in  Figure \ref{fig:profiledrop}. The profile plot shows that the flow strength is significantly reduced in the presence of confinement in case of sandwich method of protein crystallization. The flow strength at different confinement conditions has been presented. The  $u$ velocity is compared for four confinement cases. The confinement is changed by changing the height of the ceiling from the bottom cover. Details geometry of the crystallization chamber for different confinement conditions has been mentioned in the Figure \ref{fig:profiledrop}. It shows increasing confinement reduces the flow strength inside the droplet. The objective of the work is to reduce the evaporation induced flow which is the most dominant flow mechanism. The flow strength is higher at the beginning of crystallization process due to higher evaporation rate. Hence, the flow strengths for different confinement condition are compared during the initiation of the process i.e. after 10 minute of starting of the crystallization. The cause of the flow reduction is presented in the following paragraphs. 

\par
Evaporation induced convection has higher flow strength as compared to the flow strength caused by growing crystals. Evaporation induced Rayleigh convection in the pre-nucleation phase depends on the evaporation rate from the droplet surface. In our previous study \cite{Pradhan2016b}, we demonstrated that confinement by the extended wall beyond the evaporating interface suppresses the evaporation and flow strength inside aqueous droplet. In case of sitting drop method of protein crystallization, the evaporating surface remain exposed for evaporation. In case of sandwich method of protein crystallization, the liquid bridge is confined between two substrate surfaces. The extended wall of the two substrates beyond the evaporating interface suppresses the evaporation from the liquid bridge. Hence, the flow strength shows a lower value. The confinement effect of sandwich drop method helps to reduce the evaporation induced flow inside the droplet. The drop during the sitting drop method of protein crystallization is assumed to be hemispherical shape. The exposed surface area for the drop having value 2 $\mu$L is equal to \(6.09 \, \times \, 10^{-6} \, \textrm{m}^2\). Evaporation occurs from the lateral surface area of the liquid bridge in case of sandwich drop method of protein crystallization. The evaporating surface area of the liquid bridge of same volume (V=2 $\mu$L) is equal to \(3.54 \, \times \, 10^{-6} \, \textrm{m}^2\).
The evaporation from the droplet surface can be given according to Fick's law as 

\[
J=AD\frac{\partial C}{\partial n}
\]

Here, $A$ is the exposed surface area of the droplet, $D$ is the diffusive coefficient of water vapor in air and \(\frac{\partial C}{\partial n}\) is the gradient of vapor concentration at the interface of the droplet. Confinement effect due to extended channel wall reduces the gradient in vapor concentration \cite{Pradhan2016b}. Also liquid bridge has low surface area for evaporation as compared to hemispherical droplet of same volume. Hence, the evaporating surface area gets reduced in the presence of confinement which also contribute to the reduction in evaporation and flow strength inside the liquid bridge. Figure \ref{fig:profiledrop}  presents the velocity profile at four confinement conditions. Two confinement conditions are for sitting drop method of protein crystallization having different height (H) of upper ceiling from the substrate. Other two confinement conditions are for sandwich method of protein crystallization having different height of the liquid bridge. Sitting drop method of protein crystallization having higher ceiling height ($H$) from the substrate shows higher velocity strength as compared to lower ceiling height. A solid wall near to the droplet suppress the evaporation \cite{Pradhan2016b} due to confinement effect which is the case for lower ceiling height. Hence, flow strength for lower ceiling height causes comparatively less flow strength. When the droplet is sandwiched between the lower and upper surfaces, the flow strength is further suppressed due to confinement effect and less exposed surface area for evaporation. The flow strength can be further be reduced by reducing the height of liquid bridge which reduces the buoyancy force due to less vertical height. Also the confinement effect increases by reducing the liquid bridge height. It can be calculated that the exposed surface area (\(A=\pi D H\)) is directly proportional to the square root of the liquid bridge height ($H$) for the same volume (\(V=\frac{\pi}{4}D^2 H\)). 

\[
A \propto \sqrt{H}
\] 

The exposed surface area for \(H=350 \mu m\) is equal to \(2.9 \, \times \, 10^{-6} \, \textrm{m}^2\). Decreasing channel height decreases the exposed surface area leading to decrease in evaporation and flow strength inside the liquid bridge which is shown in the velocity profile.

\section{Conclusion}
In this work, we demonstrated a method to suppress the fluid flow strength using confinement during protein crystallization process. Sitting drop method of protein crystallization was used in the present study where the protein crystals are grown inside an evaporating droplet resting on a hydrophobic surface. Detail hydrodynamics study was carried out during the crystallization process.  Micro-PIV technique was used to study the fluid dynamics behavior during the process. Evaporation from the droplet induces buoyancy driven natural convection inside the droplet. Also growing crystal induces natural convection around the crystal due to concentration gradient created by the absorption of protein molecule at the crystal face. Confinement has been used to suppress the natural convection induced by evaporation. Flow strength inside the droplet was studied at four confinement conditions. With increase in confinement, flow strength decreases.  This is a very simple method and can be easily tune the evaporation induced flow strength by adjusting the confinement and the geometry of the crystal chamber.

\begin{acknowledgments}
We wish to acknowledge the Department of Science and Technology, Government of India (PROJECT NO. SR/FST/ET II-027/2006) and IIT Kanpur for the financial support to carry out the research. The research was conducted at Microfluidics and Sensor Laboratory, IIT Kanpur.
\end{acknowledgments}

\nocite{*}
\bibliography{reference}

\end{document}